\documentclass[%
 reprint,
 amsmath,amssymb,
 aps,
]{revtex4-2}

\usepackage{graphicx}
\usepackage{dcolumn}
\usepackage{bm}

\begin{document}

\preprint{APS/123-QED}

\title{Relation between the contact force and local geometry 
for an elastic curve under general surface confinement}

\author{Meng Wang}
\email{meng\_wang@pku.edu.cn}
 \affiliation{Department of Mechanics and Engineering Science, College of Engineering, Peking University, Beijing 100871, China.}

\date{\today}

\begin{abstract}
Confinement of filamentary objects is ubiquitous in numerous biological, medical, and engineering scenarios. Quantitatively determining the mechanical interaction between flexible filaments and surface confinement is particularly challenging due to the unknown contact force induced by elasticity interacting with geometric constraints. Here, we consider a simplified model of confined filamentary object: an elastic curve under surface confinement. Local force and moment balance equation incorporating the role of contact is utilized to derive the contact force exerted by surface on the confined elastic curve. It reveals the relation between contact force and local geometry at balanced state and provides a route to obtain the contact force from local confined geometry directly. Examples are provided to illustrate how to calculate contact force from obtained geometries. We believe that our results contribute to future efforts in the mechanics of filamentary objects under surface confinement. 
\end{abstract}

\maketitle


\section{\label{sec:level1}Introduction}

The slenderness of filamentary objects promises their accessibility to a narrow space, whose boundary could prevent possible penetrating trends. Such situations appear across length scales and function among natural, medical, and industrial processes. For example, DNA molecular chain is packaged inside viral capsids for storing genetic information\cite{Tao98}. Long drill string is used in petroleum engineering to drill boreholes, and could buckle inside the tube with contact with the well wall\cite{Inglis87,Wicks08}. Submillimeter-scale soft continuum robots are used to navigate through narrow space in human body for safe therapeutic and diagnostic treatment\cite{Kim19}. One physical feature shared in these situations is the geometric constraints restrict the free invasion or spread of filamentary objects, and as a result, forming contact and causing deformation in compromise.

It is difficult to theoretically determine the contact force caused by the complex mechanical interaction between filaments and boundaries over the contact region. However, the contact force between filaments and surface could have significant effects during their functioning. For example, the contact force exerted on the catheter by the wall of narrow channels is required for catheter advancement in conventional minimally invasive procedures\cite{Pancaldi20}. In petroleum industry, the helical buckling of drill pipe inside the well leads to significant contact force and decreases the access of the pipe\cite{Wicks08,Miller15}. In shaping lipid vesicle by filament, contact force transmitted by the confined actin bundles is the direct force actuating lipid membrane deformation\cite{Tsai15}, and the range of contact force may affect the diversity of deformation that the system can achieve. Similar shaping behavior appears during the interaction between cytoskeletal filaments and bacterial cells\cite{Jiang11}. These examples shed light of the demand to determine the contact force between filamentary objects and surface confinement.

As the contact behavior usually accompanied by deformation, it indicates the possibility to obtain the contact force from the deformed geometry. The contact force between a confined elastic rod and a plane or a cylinder have been derived with a parameter determined by boundary condition\cite{vanderHeijden99,vanderHeijden01}. Recently, a variational framework is performed towards a pair of interacting charged loops confined on a sphere and the contact force for trivial solution is given\cite{Chaurasia20}. For general surface confinement situation, existing theories express the contact force in terms of geometrical parameters with an additional unknow const of integration\cite{Guven12,Huynen16}, which causes malfunction in obtaining contact force from geometry. A theory that can determine the contact force from only geometry is needed to be established. Here, we derive the relation between contact force and local geometry of an elastic fiber (modeled as an elastic curve) confined on a surface from balance equation. We also consider the nonuniform bending stiffness situation which could appear in nonuniform cross section or varying material property of the confined fiber. The results for the contact force are divided into two categories according to geodesic curvature of the confined curve. For balanced confined elastic curve possessing nonvanishing geodesic curvature, the contact force is fully determined by local geometry and material properties of the elastic curve, which reveals the (local geometry)-determined property of the contact force and provides a direct approach to calculate contact force from local geometry. In contrast, for balanced confined elastic curve lies on geodesics, the contact force is shown to be determined by local geometry, material properties, and boundary condition. For the nonvanishing geodesic curvature case, the underling physical mechanism demonstrated here differs from previous work\cite{Guven12,Huynen16}, in which the unknow constant of integration is a global quantity accounting for a nonlocally determined property. In addition, due to the contact force is difficult to measure compared with local geometry, the (local geometry)-determined property promises an alternative route for measuring contact force by transmitting into measuring local geometry.
 
\section{Equations for confined problems}
Consider a flexible fiber on a general surface $\mathbf{S}$ and a perfect fiber-surface contact is assumed. The flexible fiber is modeled as an inextensible elastic curve through its centerline and the surface is parameterized as $\mathbf{S}\left ( u,v \right )$. The elastic curve on the surface can be reparameterized as $\mathbf{r}\left ( s \right )=\mathbf{S}\left ( u\left ( s \right ),v\left ( s \right ) \right )$, where $\left (u\left ( s \right ),v\left ( s \right )  \right )$ is the parameter space of the surface. As there exists no external body couple moment, the forces exert on the segment $\left [ s_{1},s_{2} \right ] \left (s_{1}< s_{2}  \right )$ are illustrated in Fig. 1, and the balance equation of the elastic fiber confined on the surface reads\cite{Audoly10}

\begin{eqnarray}
{\bf{n}}'+{\bf{f}}&=&0, \label{eq1}
\\
{\bf{m}}'+{\bf{r}}'\times {\bf{f}}&=&0, \label{eq2}
\end{eqnarray}
where $\mathbf{n}\left ( s \right )$ and $\mathbf{m}\left ( s \right )$ are the resultant internal force and moment of the stress on the cross section at $\mathbf{r}\left ( s \right )$ induced by segment $(s,s_{1}]\left ( s_{1}> s \right )$. $\mathbf{f}\left ( s \right )$ is the external body force per unit length exerting on the cross section at $s$. The prime denotes derivative with respect to the arc length $s$. The resultant internal moment $\mathbf{m}\left ( s \right )$ can be expressed by constitutive relation of elastic curve as\cite{Audoly10}
\begin{eqnarray}
\bf{m}(\mathit{s})&=&EI\mathit{\kappa} {\bf{B}}, \label{eq3}
\end{eqnarray}
where $E$ is Young’s modulus, $I$ represents moment of inertia, $\mathit{\kappa}$ denotes the curvature and $\bf{B}$ denotes unit binormal vector of the space curve. The unit binormal vector is defined in the Frenet frame \{$\bf{T}$, $\bf{N}$, $\bf{B}$\}  with the three unit vectors defined as

\begin{eqnarray}
{\bf{T}}&=&{\mathbf{r}}', \label{eq4}
\\
{\bf{N}}&=&\kappa ^{-1}{\mathbf{T}}', \label{eq5}
\\
{\bf{B}}&=&{\bf{T}}\times {\bf{N}}. \label{eq6}
\end{eqnarray}

Derivatives of these three vectors towards arc length $\mathit{s}$ are determined by the Frenet-Serret formulas as

\begin{eqnarray}
{\bf{T}}'&=&\kappa {\mathbf{N}}, \label{eq7}
\\
{\mathbf{N}}'&=& -\kappa {\mathbf{N}} + \tau {\mathbf{B}}, \label{eq8}
\\
{\mathbf{B}}'&=&-\tau {\mathbf{N}}, \label{eq9}
\end{eqnarray}
and the curvature $\mathit{\kappa}$ and torsion $\mathit{\tau}$ of the space curve can be expressed as

\begin{eqnarray}
\kappa&=& \left |{\mathbf{T}}'  \right |=\left | \mathbf{r}{}'' \right |, \label{eq10}
\\
\tau&=&{\mathbf{r}}'\cdot \left ( {\mathbf{r}}''\times {\mathbf{r}}''' \right )/\left |{\mathbf{r}}''  \right |^{2}, \label{eq11}
\end{eqnarray}
where $\left |\mathbf{x}  \right |$ denotes the norm of vector $\mathbf{x}$.

\begin{figure}[h]
\centering
\includegraphics[width=0.4\textwidth]{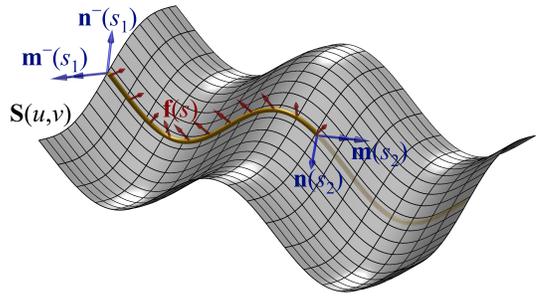}
\caption{\label{fig:epsart} Illustration of the forces acting on segment $\left [ s_{1},s_{2} \right ] \left (s_{1}< s_{2}  \right )$ of the fiber confined on a parametric surface $\mathbf{S}\left ( u,v \right )$. $\mathbf{n}^{-}\left ( s_{1} \right )$ and $\mathbf{m}^{-}\left ( s_{1} \right )$ represent the resultant internal force and moment on the cross section at $s=s_{1}$ exerted by segment $(s,s_{1}]\left ( s_{1}> s \right )$, with $\mathbf{n}^{-}\left ( s_{1} \right )=-\mathbf{n}\left ( s_{1} \right )$ and $\mathbf{m}^{-}\left ( s_{1} \right )=-\mathbf{m}\left ( s_{1} \right )$ in our convention.}
\end{figure}

We assume that there is no friction between fiber and surface, and consequently the contact force between elastic fiber and surface always directs along the normal direction of the surface at corresponding material point of the fiber. Note that there exist two normal directions for a surface and we take the opposite contact side one in this work. Thus, the contact force can be expressed as
\begin{eqnarray}
\mathbf{f}&=&-\lambda \mathbf{N}_{\mathrm{S}}. \label{eq12}
\end{eqnarray}

The unit normal vector of the surface $\mathbf{N}_{\mathrm{S}}$  can be expressed by surface parameter as
\begin{eqnarray}
\mathbf{N}_{\mathrm{S}}=\frac{\mathbf{S}_{\mathit{u}}(u,v)\times \mathbf{S}_{\mathit{v}}(u,v)}{\left | \mathbf{S}_{\mathit{u}}(u,v)\times \mathbf{S}_{\mathit{v}}(u,v) \right |}, \label{eq13}
\end{eqnarray}
where  $\mathbf{S}_{\mathit{u}}(u,v)$ and $\mathbf{S}_{\mathit{v}}(u,v)$  represent the first partial derivatives to $u$ and $v$, respectively. The unit tangent vector $\bf{T}$, unit normal vector $\mathbf{N}_{\mathrm{S}}$, together with $\mathbf{L}=\mathbf{T}\times \mathbf{N}_{\mathrm{S}}$ form a Darboux frame \{$\bf{T}$, $\mathbf{N}_{\mathrm{S}}$, $\mathbf{L}$\}, which ecolves with arc length $s$ as 
\begin{eqnarray}
{\bf{{T}'}}&=&\kappa_{n} {\mathbf{N}_{\mathrm{S}}}+\kappa _{g}\mathbf{L}, \label{eqd1}
\\
{\mathbf{N}}'_{\mathrm{S}}&=& -\kappa_{n} {\mathbf{T}} - \tau_{g} {\mathbf{L}}, \label{eqd2}
\\
{\mathbf{L}}'&=& -\kappa_{g} {\mathbf{T}} + \tau_{g} {\mathbf{N}}_{S}, \label{eqd3}
\end{eqnarray}
where $\kappa_{g}$ is the geodesic curvature, $\kappa_{n}$ is the normal curvature, and $\tau_{g}$ denotes the geodesic torsion of the curve.

Substituting Eqs. (\ref{eq3}) and (\ref{eq12}) into Eqs. (\ref{eq1})-(\ref{eq2}) and combining Eqs. (\ref{eq4})-(\ref{eq9}), the balance equation reads
\begin{eqnarray}
{\mathbf{n}}'-\lambda \mathbf{N}_{\mathrm{S}}=0, \label{eq14}
\\
{\left (EI  \right )}'\left ( s \right )\kappa \mathbf{B}+EI{\kappa }'\mathbf{B}-EI\kappa \tau \mathbf{N}+\mathbf{T}\times \mathbf{n}=0. \label{eq15}
\end{eqnarray}

Decomposing the internal force as $\mathbf{n}=n_{\mathrm{T}}\mathbf{T}+n_{\mathrm{N}}\mathbf{N}+n_{\mathrm{B}}\mathbf{B}$ and projecting the force and moment balance equation Eqs. (\ref{eq14})-(\ref{eq15}) into the Frenet frame, we can obtain

\begin{eqnarray}
{n}'_{\mathrm{T}}-\kappa n_{\mathrm{N}}&=&0, \label{eq16}
\\
{n}'_{\mathrm{N}}+\kappa n_{\mathrm{T}}-\tau n_{\mathrm{B}}-\lambda \mathbf{N}_{\mathrm{S}}\cdot \mathbf{N}&=&0, \label{eq17}
\\
{n}'_{\mathrm{B}}+\tau n_{\mathrm{N}}-\lambda \mathbf{N}_{\mathrm{S}}\cdot \mathbf{B}&=&0, \label{eq18}
\\
\left (EI  \right )\left ( s \right )\kappa \tau +n_{\mathrm{B}}&=&0, \label{eq19}
\\
{\left (EI  \right )}'\left ( s \right )\kappa +\left (EI  \right )\left ( s \right ){\kappa }'+n_{\mathrm{N}}&=&0. \label{eq20}
\end{eqnarray}

Note there exists only two independent equations in moment balance equation (\ref{eq15}), owing to the moment balance along the tangent vector $\mathbf{T}$ for elastic curve holds naturally. From the moment balance equation (\ref{eq19}) and (\ref{eq20}), we can solve $n_{\mathrm{B}}$  and $n_{\mathrm{N}}$  as 
\begin{eqnarray}
n_{\mathrm{B}}&=&-\left (EI  \right )\left ( s \right )\kappa \tau, \label{eq21}
\\
n_{\mathrm{N}}&=&-{\left (EI  \right )}'\left ( s \right )\kappa -\left (EI  \right )\left ( s \right ){\kappa }'. \label{eq22}
\end{eqnarray}

\subsection{The case of $\kappa _{g}\neq 0$}

Substituting Eqs. (\ref{eq21})-(\ref{eq22}) into Eq. (\ref{eq18}), we can obtain the contact force explicitly as

\begin{equation}
\begin{aligned}
\lambda=\frac{2{\left (EI  \right )}'\left ( s \right )\kappa^{2}\tau +2\left (EI  \right )\left ( s \right ){\kappa }'\kappa \tau +\left (EI  \right )\left ( s \right )\kappa^{2}{\tau}'}{\kappa _{g}}, \label{eq23}
\end{aligned}
\end{equation}
where we have used the relation $\mathbf{N}_{\mathrm{S}}\cdot \mathbf{B}=-\kappa _{g}/\kappa $.

The remaining internal force component $n_{\mathrm{T}}$  can be solved from Eq. (\ref{eq17}) combining with Eq. (\ref{eq23}) as
\begin{equation}
\begin{aligned}
    n_{\mathrm{T}}&={\left (EI  \right )}''\left ( s \right )+2{\left (EI  \right )}'\left ( s \right )\frac{{\kappa }'}{\kappa }+\left (EI  \right )\left ( s \right )\frac{{\kappa }''}{\kappa }-\left (EI  \right )\left ( s \right )\tau ^{2}\\
&+\frac{\kappa _{n}}{\kappa _{g}}\left (  2{\left (EI  \right )}'\left ( s \right )\tau +2\left (EI  \right )\left ( s \right )\frac{{\kappa }'}{\kappa }\tau+\left (EI  \right )\left ( s \right ){\tau}' \right ), \label{eq24}
\end{aligned}
\end{equation}
where we have used $\mathbf{N}_{\mathrm{S}}\cdot \mathbf{N}=\kappa _{n}/\kappa $.

The internal force given in Eqs. (\ref{eq21})-(\ref{eq22}) and (\ref{eq24}), and contact force given in Eq. (\ref{eq23}) can be reduced for elastic fiber possessing constant bending modulus
\begin{eqnarray}
n_{\mathrm{T}}&=&EI\left [ \frac{{\kappa}''}{\kappa }-\tau ^{2}+\frac{\kappa _{n}}{\kappa_{g} }\left ( 2\frac{{\kappa}' }{\kappa }\tau  +{\tau}'\right )   \right ], \label{eq25}
\\
n_{\mathrm{B}}&=&-EI \kappa \tau, \label{eq26}
\\
n_{\mathrm{N}}&=&-EI {\kappa}', \label{eq27}
\end{eqnarray}
and
\begin{eqnarray}
\lambda=\frac{EI \left (  2{\kappa }'\kappa \tau +\kappa^{2}{\tau}'\right )}{\kappa _{g}}. \label{eq28}
\end{eqnarray}

The contact force can also be expressed by position vector of the material point on the fiber through substituting Eqs. (\ref{eq10})-(\ref{eq11}) in to Eqs. (\ref{eq23}) and (\ref{eq28}), with the evoking of $\kappa _{g}={\mathbf{r}}''\cdot \left ( {\mathbf{r}}'\times \mathbf{N}_{S} \right )$, we can obtain
\begin{equation}
\begin{aligned}
\lambda\left ( s \right )=\left (EI  \right )\left ( s \right )\frac{{\mathbf{r}}'\cdot \left ( {\mathbf{r}}''\times {\mathbf{r}}^{\left ( 4 \right )} \right )}{{\mathbf{r}}'\cdot \left ( \mathbf{N}_{\mathrm{S}}\times {\mathbf{r}}'' \right )}+2{\left (EI  \right )}'\left ( s \right )\frac{{\mathbf{r}}'\cdot \left ( {\mathbf{r}}''\times {\mathbf{r}}^{\left ( 3 \right )} \right )}{{\mathbf{r}}'\cdot \left ( \mathbf{N}_{\mathrm{S}}\times {\mathbf{r}}'' \right )}, \label{eq29}
\end{aligned}
\end{equation}
and
\begin{eqnarray}
\frac{\lambda\left ( s \right )}{EI}=\frac{{\mathbf{r}}'\cdot \left ( {\mathbf{r}}''\times {\mathbf{r}}^{\left ( 4 \right )} \right )}{{\mathbf{r}}'\cdot \left ( \mathbf{N}_{\mathrm{S}}\times {\mathbf{r}}'' \right )}. \label{eq30}
\end{eqnarray}

\subsection{The case of $\kappa _{g}=0$}
For $\kappa _{g}=0$, from Eq. (\ref{eqd1}) we can obtain ${\mathbf{T}}'\cdot \mathbf{L}=0$. Then $\mathbf{N}=\mathbf{N}_{\mathrm{S}}$  or $\mathbf{N}=-\mathbf{N}_{\mathrm{S}}$ , and consequently $\mathbf{N}_{\mathrm{S}}\cdot \mathbf{N}=\pm 1$ . From Eq. (\ref{eq17}), we can solve
\begin{eqnarray}
\lambda=\pm  \left ( {n}'_{\mathrm{N}} +\kappa n_{\mathrm{T}}-\tau n_{\mathrm{B}}\right ). \label{eq31}
\end{eqnarray}

For $\mathbf{N}=\mathbf{N}_{\mathrm{S}}$, Eq. (\ref{eq31}) reads
\begin{eqnarray}
\lambda= {n}'_{\mathrm{N}} +\kappa n_{\mathrm{T}}-\tau n_{\mathrm{B}}, \label{eq32}
\end{eqnarray}
and we can solve $n_{\mathrm{T}}$  from Eq. (\ref{eq32}) as
\begin{eqnarray}
n_{\mathrm{T}}=\frac{\lambda-{n}'_{\mathrm{N}}+\tau n_{\mathrm{B}}}{\kappa }. \label{eq33}
\end{eqnarray}

Substituting Eq. (\ref{eq33}) into Eq. (\ref{eq16}), and combining Eqs. (\ref{eq21})-(\ref{eq22}), we can obtain a differential equation on $\lambda$ as
\begin{widetext}
\begin{equation}
\begin{aligned}
\kappa^2(E I)^{\prime}(s)+(E I)(s) \kappa \kappa^{\prime}-\tau\left(\tau(E I)^{\prime}(s)+2(E I)(s) \tau^{\prime}\right)-\frac{\kappa^{\prime}\left(\lambda +2(E I)^{\prime}(s) \kappa^{\prime}+(E I)(s) \kappa^{\prime \prime}\right)}{\kappa^2}+(EI)^{\prime \prime \prime}(s)\\
+\frac{{\lambda}'+ 2 \kappa^{\prime}(E I)^{\prime \prime}(s)
+3(E I)^{\prime}(s) \kappa^{\prime \prime}+(E I)(s) \kappa^{\prime \prime \prime}}{\kappa}=0. 
 \label{eqx}
\end{aligned}
\end{equation}
\end{widetext}

Solution to Eq. (\ref{eqx}) can be accomplished with a constant of integration $C_{1}$ as follows
\begin{widetext}
\begin{equation}
\begin{aligned}
\lambda(s)=&  \left\{C_1+\int_0^s\left[E I(u)\left(-\kappa(u) \kappa^{\prime}(u)+2 \tau(u) \tau^{\prime}(u)+\frac{\kappa^{\prime}(u) \kappa^{\prime \prime}(u)}{\kappa(u)^2}-\frac{\kappa^{\prime \prime \prime}(u)}{\kappa(u)}\right)\right.\right. \\
& \left.\left.+(E I)^{\prime}(u)\left(-\kappa(u)^2+\tau(u)^2+\frac{2 \kappa^{\prime}(u)^2}{\kappa(u)^2}-\frac{3 \kappa^{\prime \prime}(u)}{\kappa(u)}\right)-(E I)^{\prime \prime}(u) \frac{2 \kappa^{\prime}(u)}{\kappa(u)}-(E I)^{\prime \prime \prime}(u)\right] \mathrm{d} u\right\} \kappa(s).
\label{eq34}
\end{aligned}
\end{equation}
\end{widetext}

Likewise, for $\mathbf{N}=-\mathbf{N}_{\mathrm{S}}$ , we have
\begin{eqnarray}
\lambda=- \left ( {n}'_{\mathrm{N}} +\kappa n_{\mathrm{T}}-\tau n_{\mathrm{B}}\right ), 
\end{eqnarray}
from which $n_{\mathrm{T}}$ can be solved as
\begin{eqnarray}
n_{\mathrm{T}}=\frac{-\lambda-{n}'_{\mathrm{N}}+\tau n_{\mathrm{B}}}{\kappa }, \label{eq35}
\end{eqnarray}

The differential equation governing $\lambda$ then reads
\begin{widetext}
\begin{equation}
\begin{aligned}
\kappa^2(E I)^{\prime}(s)+(E I)(s) \kappa \kappa^{\prime}-\tau\left(\tau(E I)^{\prime}(s)+2(E I)(s) \tau^{\prime}\right)+\frac{\kappa^{\prime}\left(\lambda -2(E I)^{\prime}(s) \kappa^{\prime}-(E I)(s) \kappa^{\prime \prime}\right)}{\kappa^2}+(EI)^{\prime \prime \prime}(s)\\
+\frac{-{\lambda}'+ 2 \kappa^{\prime}(E I)^{\prime \prime}(s)
+3(E I)^{\prime}(s) \kappa^{\prime \prime}+(E I)(s) \kappa^{\prime \prime \prime}}{\kappa}=0, 
\nonumber
\end{aligned}
\end{equation}
\end{widetext}
and its solution gives the expression of $\lambda$  as
\begin{widetext}
\begin{equation}
\begin{aligned}
\lambda(s)=&\left\{C_2+\int_0^s\left[ E I(u)\left(\kappa(u) \kappa^{\prime}(u)-2 \tau(u) \tau^{\prime}(u)-\frac{\kappa^{\prime}(u) \kappa^{\prime \prime}(u)}{\kappa(u)^2}+\frac{\kappa^{\prime \prime \prime}(u)}{\kappa(u)}\right)\right.\right. \\
&\left.\left.+(E I)^{\prime}(u)\left(\kappa(u)^2-\tau(u)^2-\frac{2 \kappa^{\prime}(u)^2}{\kappa(u)^2}+\frac{3 \kappa^{\prime \prime}(u)}{\kappa(u)}\right)+(E I)^{\prime \prime}(u) \frac{2 \kappa^{\prime}(u)}{\kappa(u)}+(E I)^{\prime \prime \prime}(u) \right] \mathrm{d} u \right\}  \kappa(s).
\label{eq36}
\end{aligned}
\end{equation}
\end{widetext}

It shows that that the expressions of contact force $\lambda$  in Eqs. (\ref{eq34}) (for $\mathbf{N}=\mathbf{N}_{\mathrm{S}}$) and (\ref{eq36}) (for $\mathbf{N}=-\mathbf{N}_{\mathrm{S}}$) only differs in sign, consistent with the physical understanding that contact force should be equal in magnitude and opposite in direction for inner side or outer side confinement. Here, we take inner side confinement ($\mathbf{N}=-\mathbf{N}_{\mathrm{S}}$) as an example. The const of integration $C_2$ can be determined by boundary that $n_{\mathrm{T}}=F_{\mathrm{T}}$ at integration starting point $s = 0$ together with Eqs. (\ref{eq35}) and (\ref{eq36}) as
\begin{equation}
\begin{aligned}
C_2= & \left[2(E I)^{\prime}(s) \frac{\kappa^{\prime}(s)}{\kappa(s)}+(E I)^{\prime \prime}(s)\right.\\
&\left.\left.+(E I)(s)\left(\frac{\kappa^{\prime \prime}(s)}{\kappa(s)}-\tau(s)^2\right)\right]\right|_{s=0}-F_{\mathrm{T}}.
\nonumber
\end{aligned}
\end{equation}

Substituting Eqs. (\ref{eq21})-(\ref{eq22}) to Eqs. (\ref{eq18}), evoking $\mathbf{N}_{\mathrm{S}}\cdot \mathbf{B}=0$, we have the equation governing the confining geometry
\begin{equation}
2 \tau\left((E I)^{\prime}(s) \kappa+(E I)(s) \kappa^{\prime}\right)+(E I)(s) \kappa \tau^{\prime}=0. \label{eq37}
\end{equation}
For elastic curve with constant bending stiffness $(E I)^{\prime}(s)=0$, Eq. (\ref{eq37}) reduced into
\begin{equation}
2 \tau \kappa^{\prime}+\kappa \tau^{\prime}=0.
\nonumber
\end{equation}

The force and moment balance equation Eqs. (\ref{eq14}) and (\ref{eq15}) can also be projected into the fixed cartesian coordinate to solve the contact force, which reproduces Eqs. (\ref{eq29}) and (\ref{eq30}) for $\kappa _{g}\neq 0$  with the detailed procedure provided in Appendix A.

\section{Numerical examples}
To illustrate our equation, we give some examples on calculating contact force directly from local geometry of an elastic curve confined on a surface for the general case of $\kappa _{g}\neq 0$. Typical surfaces include sphere, cylinder, and torus. Here, we consider a closed curve of constant or varying bending stiffness confined on these three surfaces mentioned above.
Numerical optimization method is employed to determine the balanced state by minimizing elastic bending energy of the closed elastic curve $E_{\mathrm{el}}=\frac{1}{2} \int_0^L(E I)(s)\left|\mathbf{r}^{\prime \prime}\right|^2\mathrm{~d} s$ , where  $(E I)(s)$ is the bending stiffness of the elastic curve and $L$ denotes the total length of the curve. The parameter functions $u\left ( s \right )$ and $v\left ( s \right )$ are approximated by uniform sextic B spline curve as $u(s)=\sum_{i=0}^n a_i N_i(s)$  and $v(s)=\sum_{i=0}^n b_i N_i(s)$ , where $a_i$ and $b_i$ serve as the coefficients of basis function $N_i(s)$ defined on a uniform knot vector of $s$. We choose the uniform knot vector as $\left\{s_0, s_1, s_2, \ldots, s_{n+10}, s_{n+11}, s_{n+12}\right\}$  with $s_{i}=0$ $\left ( i=0,1,...,5,6 \right )$ and $s_{i}=1$ $\left ( i=n+6,...,n+10,n+11,n+12 \right )$, and $n$ is taken as 100 in our calculations. During optimization, the inextensible constraint  $\left|\mathbf{r}^{\prime}(s)\right|=1$ is enforced, which is equivalent to $\left(\mathbf{S}_u \cdot \mathbf{S}_u u^{\prime}(s)^2+\mathbf{S}_u \cdot \mathbf{S}_v u^{\prime}(s) v^{\prime}(s)+\mathbf{S}_v \cdot \mathbf{S}_v v^{\prime}(s)^2\right)^{1 / 2}=1$. The closure conditions $\mathbf{r}^{\left ( i \right )}\left ( 0 \right )=\mathbf{r}^{\left ( i \right )}\left ( L \right )$  $\left ( i=0,1,2,3,4 \right )$ is also enforced, where $\mathbf{r}^{\left ( i \right )}\left ( s \right )$ denotes the $i$-th order derivative towards $s$. The contact force corresponding to obtained shape calculated using Eq. (\ref{eq23}) or (\ref{eq28}) (or equivalently Eq. (\ref{eq29}) or (\ref{eq30})) together with internal force components calculated using Eqs. (\ref{eq25})-(\ref{eq27}) are depicted in Figs. 2-5.
\subsection{Spherical surface confinement}
Spherical surface can be parameterized as $\mathbf{S}(\theta, \varphi)=(R \sin \theta \cos \varphi, R \sin \theta \sin \varphi, R \cos \theta)$, and the unit normal vector reads $\mathbf{N}_{\mathrm{S}}=\frac{\mathbf{S}_\theta \times \mathbf{S}_{\varphi}}{\left|\mathbf{S}_\theta \times \mathbf{S}_{\varphi}\right|}=(\sin \theta \cos \varphi, \sin \theta \sin \varphi, \cos \theta)$. Curve lies on a sphere have the parameterization $\mathbf{r}(s)=(R \sin \theta(s) \cos \varphi(s), R \sin \theta(s) \sin \varphi(s), R \cos \theta(s))$.

\begin{figure}[h]
\centering
\includegraphics[width=0.48\textwidth]{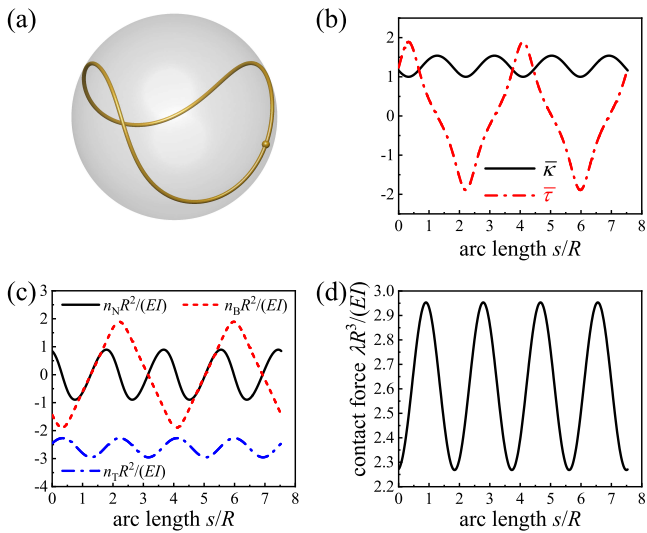}
\caption{\label{fig:epsart} Applications of Eqs. (\ref{eq25})-(\ref{eq28}) to obtain internal force and contact force for a closed elastic curve of constant bending stiffness $EI$ under spherical confinement. Numerical results for confining shape (a) and dimensionless curvature $\bar{\kappa }=\kappa R$  and dimensionless torsion $\bar{\tau  }=\tau R$  (b) of the elastic curve, the dimensionless length of the curve is $L/\left ( 2\pi R \right )=1.2$, where $L$ is the length of the curve and $R$ is the radius of the sphere. (c) the dimensionless internal force components $n_{\mathrm{T}}R^{2}/\left ( EI \right )$, $n_{\mathrm{N}}R^{2}/\left ( EI \right )$, $n_{\mathrm{B}}R^{2}/\left ( EI \right )$ calculated with the obtained geometry and Eqs. (\ref{eq25})-(\ref{eq27}), respectively. (d) dimensionless contact force $\lambda R^{3}/\left ( EI \right )$ calculated with Eq. (\ref{eq28}) and numerical geometric results. The arc length $s$ of the closed elastic curve is measured from the solid point along the positive azimuthal direction.}
\end{figure}

\subsection{Cylindrical surface confinement}
Cylindrical surface can be parameterized as $\mathbf{S}(\phi, z)=(R \cos \phi, R \sin \phi, z)$, and the unit normal vector reads $\mathbf{N}_{\mathrm{S}}=\frac{\mathbf{S}_\phi \times \mathbf{S}_z}{\left|\mathbf{S}_\phi \times \mathbf{S}_z\right|}=(\cos \phi, \sin \phi, 0)$. Curve lies on a cylinder have the parameterization $\mathbf{r}(s)=(R \cos \phi(s), R \sin \phi(s), z(s))$.

\begin{figure}[h]
\centering
\includegraphics[width=0.48\textwidth]{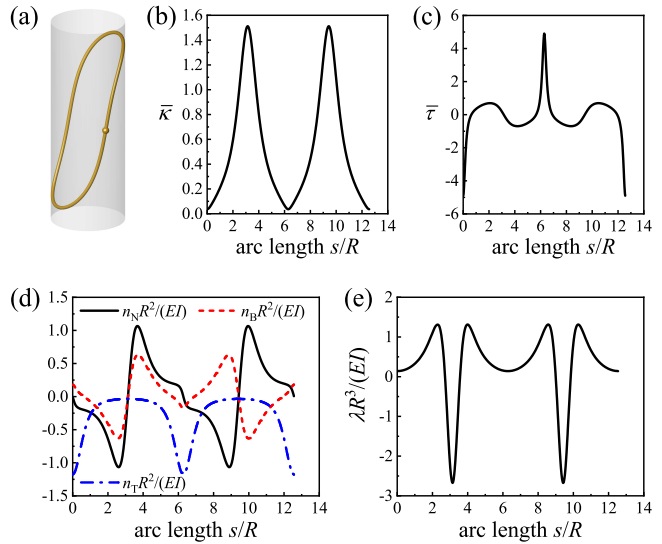}
\caption{\label{fig:epsart} Applications of Eqs. (\ref{eq25})-(\ref{eq28}) to obtain internal force and contact force for a closed elastic curve with constant bending stiffness $EI$ under cylindrical confinement. Numerical results for the shape (a), dimensionless curvature $\bar{\kappa }=\kappa R$ (b), and dimensionless torsion $\bar{\tau  }=\tau R$ (c) of the confined elastic curve with length $L=4\pi R$, where $R$ denotes the radius of the cylinder. Dimensionless internal force components $n_{\mathrm{T}}R^{2}/\left ( EI \right )$ (d), $n_{\mathrm{N}}R^{2}/\left ( EI \right )$, and $n_{\mathrm{B}}R^{2}/\left ( EI \right )$ (e) calculated with the obtained geometry and Eqs. (\ref{eq25})-(\ref{eq27}). (f) dimensionless contact force $\lambda R^{3}/\left ( EI \right )$ calculated with Eq. (\ref{eq28}) and numerical geometric results. The arc length $s$ of the closed elastic curve is measured from the solid point along the positive azimuthal direction.}
\end{figure}

\begin{figure}[h]
\centering
\includegraphics[width=0.48\textwidth]{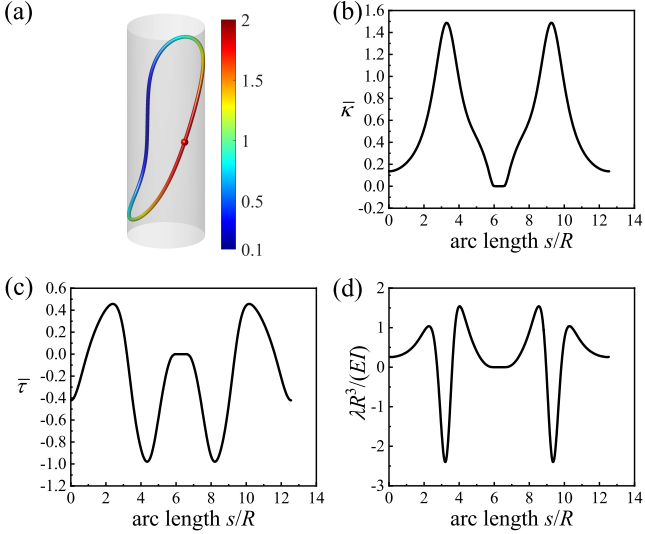}
\caption{\label{fig:epsart} Applications of Eq. (\ref{eq23}) to obtain contact force for a closed elastic curve with varying bending stiffness $[6+5 \cos (2 \pi s / L)] E I / 6$ under cylindrical confinement. Numerical results for confining shape (the rescaled bending stiffness $[6+5 \cos (2 \pi s / L)] / 6$ is encoded by the color bar) (a), dimensionless curvature $\bar{\kappa }=\kappa R$ (b) and dimensionless torsion $\bar{\tau  }=\tau R$ (c) of the elastic curve. The length of the curve is $L=4\pi R$, where $R$ is the radius of the cylinder. (d) dimensionless contact force $\lambda R^{3}/\left ( EI \right )$) calculated with Eq. (\ref{eq23}) and numerical geometric results. The arc length $s$ of the closed elastic curve is measured from the solid point along the positive azimuthal direction.}
\end{figure}

\subsection{Torus confinement}
The surface of a torus can be parameterized as $\mathbf{S}(\theta, \varphi)=(R+a \cos \phi) \cos \theta,(R+a \cos \phi) \sin \theta, a \sin \phi)$, and the unit normal vector reads $\mathbf{N}_{\mathrm{S}}=\frac{\mathbf{S}_\theta \times \mathbf{S}_\phi}{\left|\mathbf{S}_\theta \times \mathbf{S}_\phi\right|}=(\cos \phi \cos \theta, \cos \phi \sin \theta, \sin \phi)$. Curve lies on a cylinder have the parameterization $\mathbf{r}(s)=(R+a \cos \phi(s)) \cos \theta(s),(R+a \cos \phi(s)) \sin \theta(s), a \sin \phi(s))$.

\begin{figure*}[htb]
\includegraphics[width=0.9\textwidth]{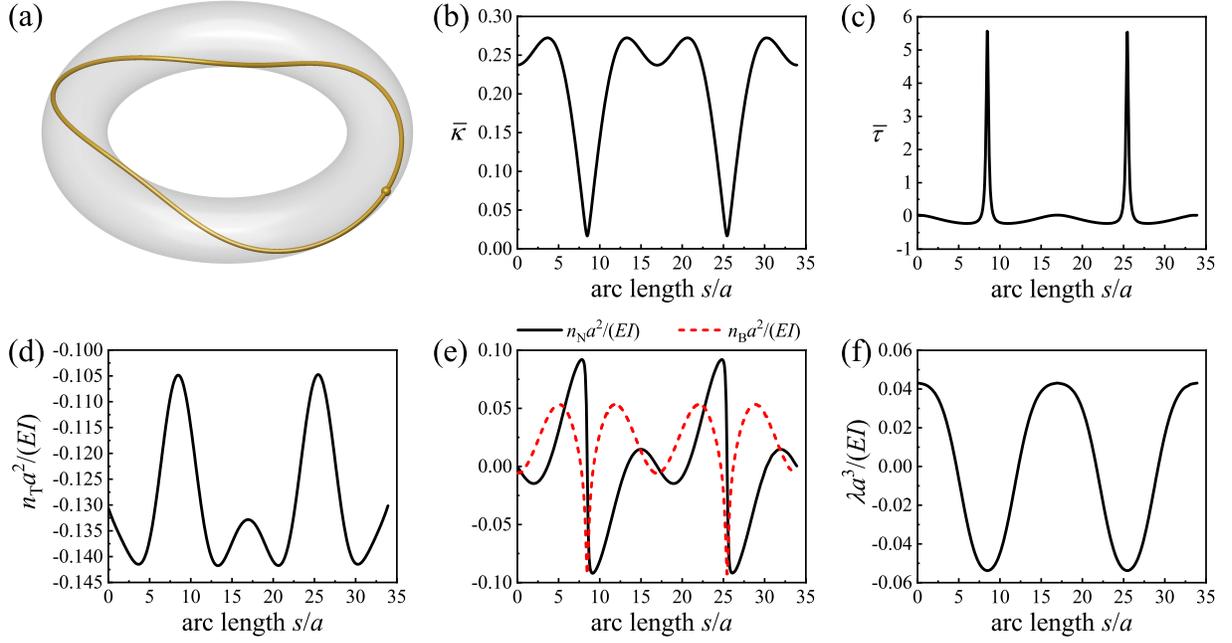}
\caption{\label{fig:epsart} Applications of Eqs. (\ref{eq25})-(\ref{eq28}) to obtain internal force and contact force for a closed elastic curve with constant bending stiffness $EI$ confined on the surface of a torus. Numerical results for confining shape (a) and dimensionless curvature $\bar{\kappa }=\kappa R$  (b) and dimensionless torsion $\bar{\tau  }=\tau R$ (c) of the elastic curve. The length of the curve is $L=1.8\pi \left ( R+a \right )$, where $R$ is the major radius and $a$ is the minor radius of the torus with $R=5a$ taken here. Dimensionless internal force components $n_{\mathrm{T}}a^{2}/\left ( EI \right )$ (d), $n_{\mathrm{N}}a^{2}/\left ( EI \right )$, and $n_{\mathrm{B}}a^{2}/\left ( EI \right )$ (e) calculated with the obtained geometry and Eqs. (\ref{eq25})-(\ref{eq27}). (f) dimensionless contact force $\lambda R^{3}/\left ( EI \right )$ calculated with Eq. (\ref{eq28}) and numerical geometric results. The arc length $s$ of the curve is measured from the solid point along the positive azimuthal direction.}
\end{figure*}

\section{Discussion}

In Eqs. (\ref{eq23}) and (\ref{eq28})-(\ref{eq30}), forth order derivative of position vector $\mathbf{r}\left ( s \right )$ requires up to $C^{4}$ continuous, and confining surface should be continuous to calculate the normal vector. Besides, at least $C^{1}$ continuous is required for bending rigidity $EI\left ( s \right )$ in Eqs. (\ref{eq23}) and (\ref{eq29}). As our results are derived from balance equation, the balanced shape is required for calculating contact force. On the other hand, the equations can also give the force matching a configuration on a surface to reach a balanced state. Due to the balance equation used in derivation focuses on local segment without concerning outer parts for the case of non-vanishing geodesic curvature, whether the outer parts lying on the surface or detaching from the surface do not affect our results for the contact region we examined. The effects of outer region geometry may enter in local geometry through global balance condition of the whole constrained fiber, and so as the effects of boundary condition. Then local geometry and material properties can fully capture the contact force without nonlocal properties involved for the case of non-vanishing geodesic curvature. As there is no boundary assumption in our derivation, our equation can be applied to both closed curve and open curve. There is no scale assumption in our equation, which promises our theory to hold across length scales whenever the elasticity dominates. No assumption on the deformability of the confined surface is made. Accordingly, our results can be applied to elastic curve confined on both rigid and deformable surfaces.

Further extension of our study can be focused on magnetic-actuated slender objects, relevant mechanical models of slender hard-magnetic soft materials have been proposed recently\cite{Zhao19,Sano22}. The fast and programmable response feature of magnetic-actuated slender objects support their broad prospects for working under surface confinement, and the contact force imposed by the confinement could be expected to take the magnetic reaction into account.

\appendix

\section{Solving contact force from fixed cartesian coordinate}
Here, we provide an alternative derivation for solving the contact force through projecting the governing equations into a fixed frame $Oxyz$.
Substituting Eqs. (\ref{eq3}) and (\ref{eq12}) into Eqs. (\ref{eq1}) and (\ref{eq2}), and combining Eqs. (\ref{eq4})-(\ref{eq9}), the balance equation reads
\begin{eqnarray}
{\mathbf{n}}'-\lambda \mathbf{N}_{\mathrm{S}}=0, \label{eqa1}
\\
{\left (EI\left ( s \right ){\mathbf{r}}'\times {\mathbf{r}}''  \right )}'+{\mathbf{r}}'\times \mathbf{n}=0. \label{eqa2}
\end{eqnarray}

We denote ${\left (EI\left ( s \right ){\mathbf{r}}'\times {\mathbf{r}}''  \right )}'=\left ( A_{x},A_{y},A_{z} \right )$, ${\mathbf{r}}'=\left ( B_{x},B_{y},B_{z} \right )$, $\mathbf{N}_{\mathrm{S}}=\left ( N_{\mathrm{S}}^{x},N_{\mathrm{S}}^{y},N_{\mathrm{S}}^{z} \right )$, and $\mathbf{n}=\left ( n_{x},n_{y},n_{z} \right )$. Projecting the force and moment balance equation Eqs. (\ref{eqa1}) and (\ref{eqa2}) into the fixed $Ox$, $Oy$, $Oz$ axis of cartesian coordinate, we can obtain force balance equation (\ref{eqa1}) in its componential form as
\begin{eqnarray}
n_x^{\prime}(s)-\lambda(s) N_{\mathrm{S}}^x(s)&=&0, \label{eqa3}
\\
n_y^{\prime}(s)-\lambda(s) N_{\mathrm{S}}^y(s)&=&0, \label{eqa4}
\\
n_z^{\prime}(s)-\lambda(s) N_{\mathrm{S}}^z(s)&=&0, \label{eqa5}
\end{eqnarray}
and the moment balance equation Eq. (\ref{eqa2}) as
\begin{eqnarray}
A_x(s)-B_z(s) n_y(s)+B_y(s) n_z(s)&=&0, \label{eqa6}
\\
A_y(s)+B_z(s) n_x(s)-B_x(s) n_z(s)&=&0, \label{eqa7}
\\
A_z(s)-B_y(s) n_x(s)+B_x(s) n_y(s)&=&0. \label{eqa8}
\end{eqnarray}

From the moment balance equation (\ref{eqa8})  and (\ref{eqa6}) , we can solve $n_x(s)$ and $n_z(s)$ in expression of $n_y(s)$ as
\begin{eqnarray}
n_x(s)&=&\frac{A_z(s)+B_x(s) n_y(s)}{B_y(s)}, \label{eqa9}
\\
n_z(s)&=&\frac{-A_x(s)+B_z(s) n_y(s)}{B_y(s)}. \label{eqa10}
\end{eqnarray}

From Eq. (\ref{eqa4}), we can obtain

\begin{equation}
n_y^{\prime}=\lambda(s) N_{\mathrm{S}}^y(s).\label{eqa11}
\end{equation}

Differential Eq. (\ref{eqa9}) with $s$ and combining Eqs. (\ref{eqa3}) and (\ref{eqa11}), we can solve $n_y(s)$ as
\begin{equation}
\begin{aligned}
n_y(s)=&\frac{-A_z(s) B_y^{\prime}(s)+A_z^{\prime}(s) B_y(s)}{B_x(s) B_y^{\prime}(s)-B_x^{\prime}(s) B_y(s)}\\
&+\frac{\lambda(s) B_y(s)\left(B_x(s) N_s^y(s)-B_y(s) N_s^x(s)\right)}{B_x(s) B_y^{\prime}(s)-B_x^{\prime}(s) B_y(s)}. \label{eqa12}
\end{aligned}
\end{equation}

Substituting Eq. (\ref{eqa12}) into Eq. (\ref{eqa4}), we can solve $\lambda(s)$ with one constant of integration. Then, combining Eqs. (\ref{eqa12}), (\ref{eqa10}) and (\ref{eqa5}), the const of integral can be obtained and substitute into expression of $\lambda(s)$ to eliminate the const, we have the expression of $\lambda(s)$ as
\begin{widetext}
\begin{equation}
\begin{aligned}
\lambda(s)=-\frac{B_y^{\prime 2}\left(A_x B_x+A_z B_z\right)-B_y(s) B_y^{\prime}\left(A_x B_x+A_z B_z\right)^{\prime}+B_y^2\left(A_x^{\prime} B_x^{\prime}+A_z^{\prime} B_z^{\prime}\right)}{B_y^2\left[B_z\left(N_{\mathrm{S}}^x B_y^{\prime}-B_x^{\prime} N_{\mathrm{S}}^y\right)+B_y\left(B_x^{\prime} N_{\mathrm{S}}^z-N_S^x B_z^{\prime}\right)+B_x\left(N_{\mathrm{S}}^y B_z^{\prime}-B_y^{\prime} N_{\mathrm{S}}^z\right)\right]},
\nonumber
\end{aligned}
\end{equation}
\end{widetext}
further simplification through recovering previous notes leads to
\begin{equation}
\begin{aligned}
\lambda\left ( s \right )=\left (EI  \right )\left ( s \right )\frac{{\mathbf{r}}'\cdot \left ( {\mathbf{r}}''\times {\mathbf{r}}^{\left ( 4 \right )} \right )}{{\mathbf{r}}'\cdot \left ( \mathbf{N}_{\mathrm{S}}\times {\mathbf{r}}'' \right )}+2{\left (EI  \right )}'\left ( s \right )\frac{{\mathbf{r}}'\cdot \left ( {\mathbf{r}}''\times {\mathbf{r}}^{\left ( 3 \right )} \right )}{{\mathbf{r}}'\cdot \left ( \mathbf{N}_{\mathrm{S}}\times {\mathbf{r}}'' \right )}. \label{eqa13}
\end{aligned}
\end{equation}

For elastic fiber possessing constant bending modulus, Eq. (\ref{eqa13}) can be reduced into
\begin{eqnarray}
\frac{\lambda\left ( s \right )}{EI}=\frac{{\mathbf{r}}'\cdot \left ( {\mathbf{r}}''\times {\mathbf{r}}^{\left ( 4 \right )} \right )}{{\mathbf{r}}'\cdot \left ( \mathbf{N}_{\mathrm{S}}\times {\mathbf{r}}'' \right )}.
\nonumber
\end{eqnarray}

\nocite{*}

\bibliography{apssamp}

\end{document}